\documentclass[sigconf]{acmart}

\usepackage{mathtools}
\usepackage[capitalize,noabbrev]{cleveref}

\AtBeginDocument{%
  }

\setcopyright{none}
\renewcommand\footnotetextcopyrightpermission[1]{}
\acmConference[AgenticSE @ KDD 2026]{KDD 2026 Workshop on Agentic Software Engineering}{August 9, 2026}{Jeju, Korea}
\acmYear{2026}
\copyrightyear{2026}
\acmDOI{}
\acmISBN{}

\acmISBN{}

\begin{document}

\title{Memory as a Service (MaaS): Purpose-Bound Memory Mediation for Cooperative Agents}

\author{Haichang Li}
\email{hli52@gmu.edu}
\orcid{0009-0006-0952-0709}
\affiliation{%
  \institution{George Mason University}
  \city{Fairfax}
  \state{VA}
  \country{USA}
}

\renewcommand{\shortauthors}{Haichang Li}

\begin{abstract}
  Agentic programming is code-centered, while its useful memory context extends beyond code. A programming agent may draw on memory from test, review, build, and release agents; design, product, security, operations, and compliance agents; meeting, finance, calendar, and workflow agents; personal agents; and agents acting for other people. These memories can help agents optimize, debug, test, and evaluate software, while carrying different owners, purposes, recipients, and disclosure boundaries. We propose \emph{Memory as a Service} (MaaS) as \emph{purpose-bound memory mediation}: each invocation is evaluated by owner, requester, recipient, task, and declared purpose, and the mediator chooses whether to \emph{withhold}, \emph{abstract}, or \emph{reveal} each candidate item. We formalize this by separating cooperative utility, disclosure leakage, and purpose-bound authorization, then ground the position with diagnostic stress tests on MAGPIE. Relevance-based retrieval reaches AUROC $0.570$ and leaks $53.0\%$ of private items; contextual-integrity prompting reduces leakage by $21.8$ percentage points while leaving $32.6\%$ residual leakage; and $4.5\%$ of private items contain explicit safe-hint abstractions. These probes motivate memory governance as a separate design problem for cooperative programming agents.
\end{abstract}




\received[accepted]{10 June 2026}

\maketitle

\section{Introduction}
\label{sec:intro}

Agentic programming is code-centered, while useful memory extends beyond code. Programming agents call tools, keep state, revise outputs, and accumulate memory from test, debug, review, build, release, design, product, security, operations, and compliance agents. Context may also come from meeting, finance, calendar, workflow, customer-facing, personal, reviewer, manager, designer, and customer agents, shaping how code is optimized, tested, explained, and released.

Consider an enterprise meeting-minutes system. A programming agent improving it may need evidence from meeting and workflow agents: missed action items, wrong workflow classifications, meeting failures, or~sensitive discussion summarized at the wrong granularity. This evidence can guide model, prompt, evaluation, and process changes while carrying strategy, personnel discussion, customer information, financial plans, legal concerns, and security incidents. Useful programming and sensitive organizational context can coexist in one remembered item.

Protocols and agent frameworks increasingly support tool use, connection, and agent-to-agent interaction \cite{anthropic2024mcp,google2025a2a,wu2024autogen,hong2024metagpt}. The design question is which remembered context may be used, by whom, for whom, in which form, and for which purpose. A programming agent may need ``action-item extraction fails when ownership is implied across turns,'' while raw meeting content stays inside the owning memory service. A release agent may need ``this module needs stabilization for enterprise workloads,'' while customer names and incident blame stay there. The memory interface needs usefulness, authorization, recipient, and disclosure form.

The issue spans many owners and policies. Project histories, policies, meeting records, shared experience, personal preferences, and other people's agent memories can support software work across individual, customer, and organizational boundaries. The recurring problem is the same: \emph{the unit that stores memory is distinct from the unit that needs to invoke it}. This paper argues for \emph{Memory as a Service} (MaaS): contextual memory as a governed service invoked for a declared purpose. A request specifies owner, requester, recipient, task, purpose, and context; for each item, a mediator chooses \textsf{withhold}, \textsf{abstract}, or \textsf{reveal}. This vocabulary puts purpose, recipient, authorization, and disclosure granularity into the memory interface itself.

This position paper makes the problem precise for cooperative programming agents through purpose-bound memory mediation. It separates relevance, authorization, and leakage, then uses MAGPIE as a diagnostic probe. Relevance weakly predicts shareability (AUROC $0.570$), and top-$k$ retrieval leaks $53.0\%$ private memory while recovering useful context. These results motivate purpose-aware authorization, policy metadata, abstraction, audit, and compositional-leakage work.

\section{Related Work}
\label{sec:related}

\paragraph{Agent memory.} LLM memory systems make persistence practical through external memory management, long-term user memory, memory editing, graphs, and reusable workflows \cite{packer2024memgpt,zhong2024memorybank,xu2025amem,chhikara2025mem0,memgraph2025,wang2024agentworkflow}. Shared or operating-system-like memory layers loosen agent-local storage \cite{mem02025openmemory,li2025memosoperatingmemoryaugmentedgeneration}, while memory sharing and collaborative memory study pooled or multi-user stores \cite{gao2024memorysharing,rezazadeh2025collaborative}. Personal-agent systems show the same pressure in deployed settings, where persistent memory and social or group interaction coexist \cite{steinberger2025openclaw,nousresearch2026hermes,wei2025secondme,naturalselection2026elys,youware2026bloome,moltbook2026,moonshot2026kimi}. MaaS focuses on the policy surface that appears once a memory item is callable across an owner, agent, or trust boundary.

\paragraph{Agent communication, access, and privacy.} Multi-agent frameworks and protocols coordinate agents, tools, and agent-to-agent communication \cite{wu2024autogen,hong2024metagpt,anthropic2024mcp,google2025a2a}; identity work studies authentication, delegation, and accountability \cite{levi2026aip}. These mechanisms answer who can connect and how messages move. Memory invocation adds a content-level decision: whether a remembered detail should be withheld, abstracted, or revealed for the declared task. Security work on long-term memory identifies poisoning, unintended sharing, and cross-user contamination \cite{lin2026surveysecuritylongtermmemory,lam2026governing}. Contextual integrity supplies the relevant flow parameters--sender, recipient, information type, purpose, and transmission principle \cite{nissenbaum2009privacy}; recent LLM privacy work and benchmarks study related leakage and checking problems \cite{shao2024privacylens,li-etal-2025-1,cheng2026privact,mireshghallah2023confaide,shao2025privacylens}. MaaS places those questions at the boundary where owned memory becomes a service for cooperative software-engineering agents.

\section{Purpose-Bound Memory Mediation}
\label{sec:framework}

\subsection{Core position}

We frame MaaS as purpose-bound memory mediation: a service layer governing how contextual memory is invoked, abstracted, or disclosed across cooperative agents. A memory module should be independently addressable, composable with other memory modules, and governable by intent, so the decision depends on why memory is requested as well as who requests it.

This is a design position rather than a finished architecture. It moves the question from "does this agent have access to that memory?" to "what form of memory invocation is authorized for this purpose, requester, and recipient?"

\subsection{Private memory with a service interface}

MaaS starts from memory as a private asset. Contextual memory records lived experience, preferences, constraints, or task history, and may be owned by a person, agent, team, or organization. NIST's definition of a data asset, ``any entity that is comprised of data can be considered an asset'' \cite{nist2025dataasset}, gives a useful baseline. Ownership, control, and governance should stay attached to the memory module.

Cooperation still requires a public-facing interface. A private memory may need to answer another agent's request through a bounded service with scope, duration, recipient, purpose, and revocation conditions, while the memory store remains owned and governed.

\subsection{Memory invocation and mediation}

Consider $N$ agents $\mathcal{A} = \{A_1, \ldots, A_N\}$, each with a contextual memory store $M_i$. When a cooperative task needs information held by another agent, the requester issues a memory invocation:
\begin{equation}
q = (\underbrace{o}_{\text{owner}}, \underbrace{r}_{\text{requester}}, \underbrace{c}_{\text{recipient}}, \underbrace{t}_{\text{task}}, \underbrace{p}_{\text{purpose}}, \underbrace{x}_{\text{context}}) \in \mathcal{Q},
\label{eq:request}
\end{equation}
where $o$ is the memory owner, $r$ the requester, $c$ the intended recipient, $t$ the task, $p$ the declared purpose, and $x$ additional context. The tuple records what is sought, who asks, who receives the answer, and why. This follows contextual integrity, where information flow is evaluated through contextual parameters rather than content alone \cite{nissenbaum2009privacy}.

Given $q$, a mediator selects an action for each candidate memory item $m \in M_o$:
\begin{equation}
\pi(q, m) \in \{\mathsf{withhold},\, \mathsf{abstract},\, \mathsf{reveal}\}.
\label{eq:action}
\end{equation}
\textsf{withhold} returns no content. \textsf{reveal} returns the item, or a close paraphrase. \textsf{abstract} returns a task-sufficient statement with some detail removed, such as ``requires stabilization work'' instead of crash frequency, private customer names, or internal blame. This middle action matters because binary access models force a choice between silence and raw disclosure.

\subsection{Owner-defined access policies}

Purpose-bound mediation needs memory-holder policies. We use two layers.

\textbf{Structured constraints.} Some rules precede relevance or utility. A user may mark medical records as unavailable to agents outside healthcare, or salary information as unavailable to recruiting agents. If item $m$ falls under such a rule and $q$ matches the restricted recipient or purpose, the mediator removes \textsf{reveal} from the action set, or selects \textsf{withhold} directly.

\textbf{Policy authoring instructions.} Many owners will express disclosure preferences as ordinary instructions, such as ``keep my project experience at Company X away from recruiting agents'' or ``vacation plans may be shared with family agents and kept away from work colleagues.'' In MaaS, such instructions are an authoring interface; enforcement remains structured conditions over information type, recipient class, purpose, expiry, and permitted action. Mapping ordinary instructions into those conditions is a MaaS research problem, especially when an ambiguous request requires clarification, conservative action selection, or uncertainty recorded for audit.

Both revocable layers feed into the authorization function below. Structured constraints give crisp boundaries. Authoring instructions let the owner express contextual intent in ordinary language.

\subsection{Mediation objective}

Let $U(a,m,q) \in \mathbb{R}$ denote cooperative task utility, $L(a,m,q) \in \mathbb{R}_{\geq 0}$ denote disclosure leakage, and $A(m,q,a)\in\{0,1\}$ denote purpose-bound authorization. A simple mediation objective is
\begin{equation}
\max_{a} \, U(a, m, q) - \lambda L(a, m, q) \quad \text{s.t.} \quad A(m, q, a) = 1,
\label{eq:objective}
\end{equation}
where $\lambda \geq 0$. The equation is a modeling device, not a claim that MaaS must be implemented as an optimizer. It separates three quantities retrieval systems often collapse into one score: relevance as utility evidence, authorization as policy judgment, and leakage as disclosure cost. A highly relevant item can still be unauthorized; an authorized item can still require abstraction.

\subsection{Two structural claims}

Let $s(m,q)\in\mathbb{R}$ be a relevance score. A relevance-only rule selects \textsf{reveal} if $s(m,q)\geq\theta$, and \textsf{withhold} otherwise. Let $S$ and $P$ denote the score distributions for shareable and private items. In practice, these distributions may overlap. In that region, private and shareable memories can receive similar scores.

\begin{proposition}[Relevance is an incomplete policy signal]
\label{prop:auth}
If $S$ and $P$ overlap on a set with nonzero probability mass, then any threshold rule on $s(m,q)$ that retrieves shareable items from that overlap also retrieves private items from that overlap.
\end{proposition}

\noindent The proof follows directly from overlap. A score interval containing both shareable and private items makes any threshold that includes shareable items in that interval also include private items. The claim is modest: relevance is useful evidence, but not an authorization rule.

\begin{proposition}[Abstraction can dominate binary access]
\label{prop:abstract}
For a request $q$ and item $m$, suppose \textsf{reveal} has useful task content but leaks detail unnecessary for $q$. If an abstraction preserves enough task utility and removes that unnecessary detail, then \textsf{abstract} is preferable to both \textsf{withhold} and \textsf{reveal} for that request.
\end{proposition}

\noindent In Equation~\ref{eq:objective}, this means abstraction has higher utility than withholding and lower leakage than revealing. The condition identifies cases where cooperation needs some information, while direct disclosure gives too much. The MAGPIE safe-hint items in Section~\ref{sec:empirical} are examples.

\begin{figure}[ht]
\vskip -0.1in
\begin{center}
\includegraphics[width=\linewidth]{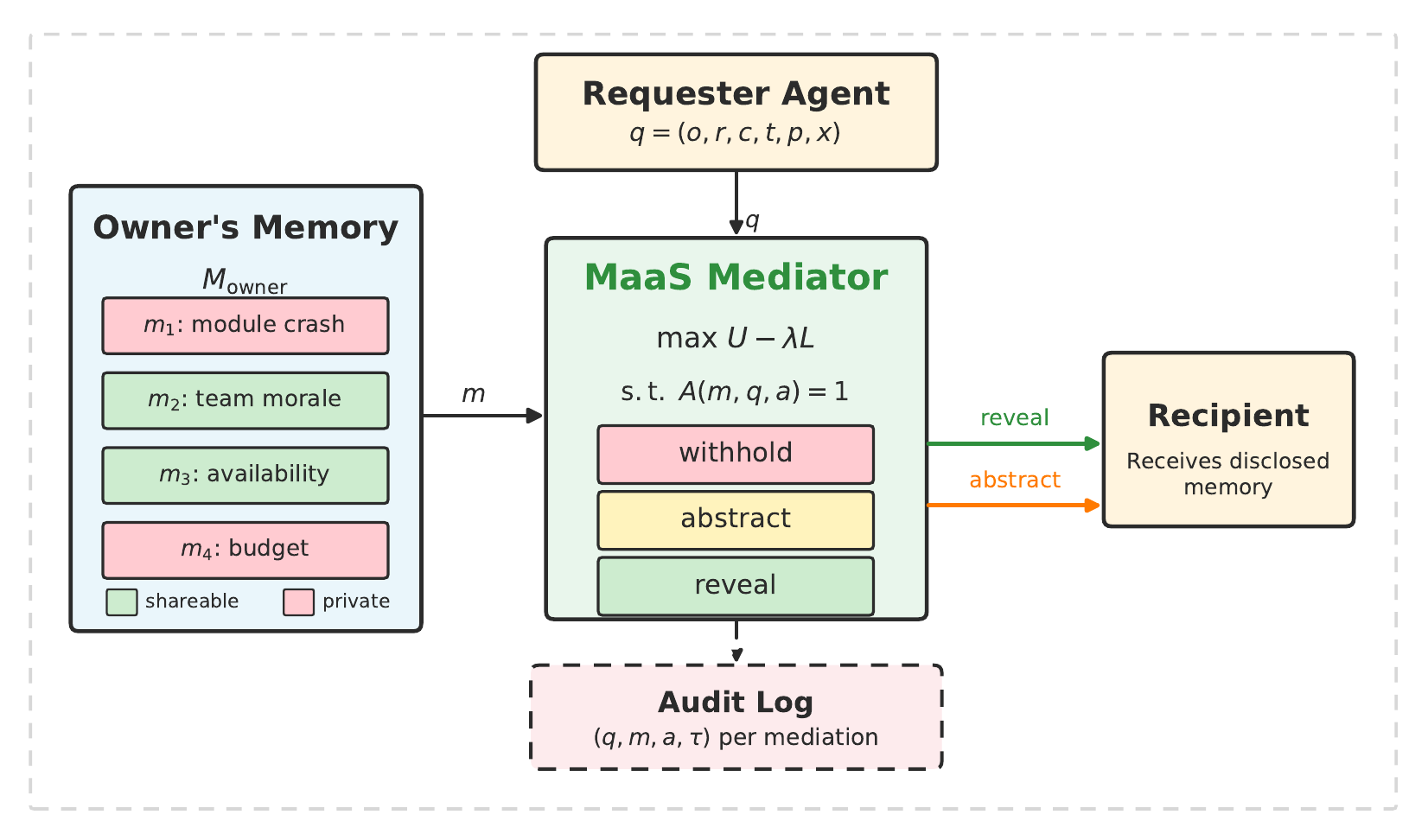}
\caption{Purpose-bound memory mediation. A request $q$ and memory item $m$ enter a mediator. The mediator selects \textsf{withhold}, \textsf{abstract}, or \textsf{reveal}, then records the decision for audit.}
\Description{Diagram of a memory mediation pipeline where a request and memory item enter a mediator that chooses to withhold, abstract, or reveal content and logs the decision for audit.}
\label{fig:pipeline}
\end{center}
\vskip -0.2in
\end{figure}

\section{Empirical Grounding: Diagnostic Stress Tests}
\label{sec:empirical}

The experiments below are diagnostic probes, not a ranking of MaaS implementations or optimal disclosure policies. They test three assumptions that would make MaaS unnecessary: relevance can stand in for authorization, a generic safety framing is enough, and binary access is expressive enough.

\subsection{Dataset and framing}

We use MAGPIE \cite{juneja2025magpie}, a public multi-agent privacy benchmark. After preprocessing, the probe set contains 200 cooperative scenarios, 1,019 agent-level probes, and 3,730 memory-like items: 2,274 shareable and 1,456 private. MAGPIE also includes safe-hint annotations for some private content. It is not representative of all MaaS settings, but has the ingredients needed here: agents, cooperative tasks, private facts, and safe abstractions.

\subsection{Stress test 1: relevance as authorization}

\textbf{Setup.} We score each memory item by cosine similarity between the item text and task-plus-deliverable query using \texttt{text-\allowbreak embedding-\allowbreak 3-\allowbreak small}, across all 1,019 probes.

\textbf{Results.} Table~\ref{tab:relevance} reports AUROC 0.570, average precision 0.659, 53.0\% private leak rate, and 41.2\% rank inversion. Relevance weakly separates shareable from private items in this dataset, matching Proposition~\ref{prop:auth}: useful and private memories often occupy similar relevance ranges.

\begin{table}[ht]
\caption{Relevance separability.}
\label{tab:relevance}
\vskip 0.02in
\centering
\begin{small}
\setlength{\tabcolsep}{4.5pt}
\begin{tabular}{lclc}
\toprule
\textbf{Metric} & \textbf{Value} & \textbf{Metric} & \textbf{Value} \\
\midrule
AUROC & 0.570 & Avg.~Precision & 0.659 \\
Private Leak & 53.0\% & Shareable Recall & 67.6\% \\
Rank Inversion & 41.2\% & & \\
\bottomrule
\end{tabular}
\end{small}
\vskip -0.15in
\end{table}

\subsection{Stress test 2: natural-language prompts as governance}

\textbf{Setup.} We sample 300 probes with seed 20260509 and evaluate four prompt conditions: \textbf{Helpful} for default cooperative behavior, \textbf{Safe} for a general safety instruction, \textbf{PurposeBound} for explicit purpose framing, and \textbf{ContextualIntegrity} for structured information-flow norms \cite{nissenbaum2009privacy}. The model and prompts probe framing sensitivity, not proposed MaaS policies.

\begin{table}[ht]
\caption{Prompt stress-test results. Intervals are 95\% bootstrap CIs from 2,000 resamples.}
\label{tab:prompt}
\vskip 0.1in
\begin{center}
\begin{small}
\begin{tabular}{lcccc}
\toprule
\textbf{Condition} & \textbf{S-Rec.} & \textbf{Prec.} & \textbf{P-Leak} & \textbf{B-F1} \\
\midrule
Helpful & 0.580 & 0.670 & 0.544 & 0.392 \\
& \scriptsize[.566,.597] & \scriptsize[.643,.695] & \scriptsize[.496,.595] & \scriptsize[.351,.435] \\[2pt]
Safe & 0.609 & 0.718 & 0.466 & 0.457 \\
& \scriptsize[.589,.629] & \scriptsize[.690,.745] & \scriptsize[.413,.517] & \scriptsize[.412,.501] \\[2pt]
PurposeBound & 0.563 & 0.708 & 0.479 & 0.414 \\
& \scriptsize[.546,.582] & \scriptsize[.680,.737] & \scriptsize[.429,.531] & \scriptsize[.373,.453] \\[2pt]
\textbf{ContextualI.} & \textbf{0.618} & \textbf{0.795} & \textbf{0.326} & \textbf{0.555} \\
& \scriptsize[.598,.639] & \scriptsize[.765,.823] & \scriptsize[.281,.376] & \scriptsize[.517,.594] \\
\bottomrule
\end{tabular}
\end{small}
\end{center}
\vskip -0.15in
\end{table}

\textbf{Results.} ContextualIntegrity reduces leakage by 21.8 percentage points relative to Helpful ($\Delta=-0.218$ [\textminus0.263, \textminus0.174]) and improves balanced-F1 by 16.3 points ($\Delta=+0.163$ [0.129, 0.199]). Residual leakage remains 32.6\% in this prompt-only setup. The result shows that purpose and recipient framing measurably affect disclosure decisions, while stronger gates remain necessary.

\subsection{Stress test 3: binary access as the action space}

\textbf{Setup.} We search the 1,456 private items for explicit safe-hint language, including ``can share that [X]'', ``can hint at [X]'', and ``can only share that [X]''.

\textbf{Results.} We find 66 such items, or 4.5\% of private items, a lower bound because the regex detects only explicit safe hints. Examples include ``needs stabilization work'' without crash details, ``team morale issues'' without attrition risk, and ``upper management pressure'' without a direct threat. These examples show why \textsf{abstract} belongs in the action space: some private items are unsafe as raw disclosures but useful as narrow summaries.

\begin{figure}[ht]
\vskip -0.1in
\begin{center}
\includegraphics[width=\linewidth]{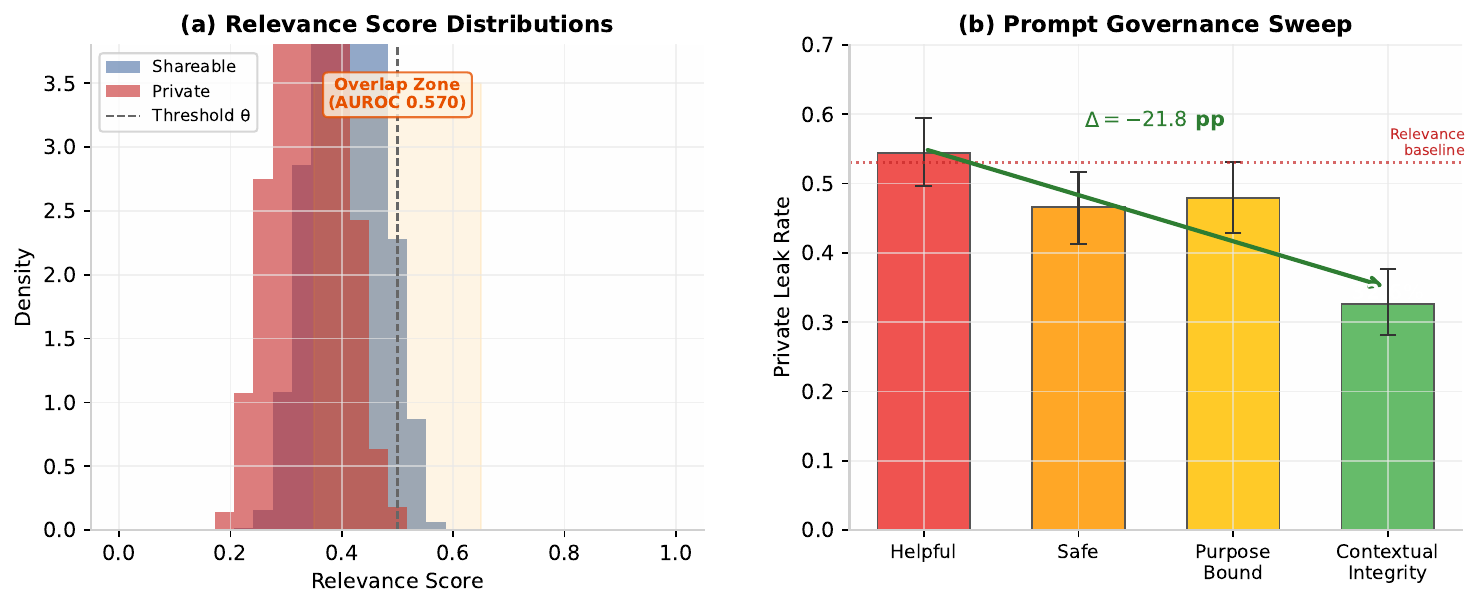}
\caption{Diagnostic results on MAGPIE. Relevance weakly separates shareable from private memory, prompt framing changes disclosure behavior, and safe hints show why binary access is too coarse for some memory items.}
\Description{Summary plots showing weak relevance separation between shareable and private memory, reduced leakage under contextual-integrity prompting, and examples motivating abstraction.}
\label{fig:results}
\end{center}
\vskip -0.2in
\end{figure}

\section{Discussion}
\label{sec:discussion}

The probes suggest a research agenda for MaaS. The first problem is policy expression: rules over owner, requester, recipient, task, purpose, time, and information type, such as ``share the project deadline with a planning agent,'' ``hide client escalation history from a recruiting agent,'' or ``allow this summary for the next incident meeting only.'' Contextual integrity views privacy as appropriate information flow, evaluated through actors, information type, purpose, and transmission principle \cite{nissenbaum2009privacy}. MaaS needs a system version of that vocabulary, attached to memory items, carried across invocations, and evaluated at request time.

This policy problem is harder than ordinary access control because the permitted action may depend on answer form. A rule may allow a release date while disallowing the delay reason, or allow disclosure to an incident-response agent until the incident closes. Policy authoring instructions make the problem realistic: teams will write guidance such as ``summarize this dependency risk for planning, excluding customer names,'' while the system maps that text into structured policy conditions. A MaaS system must translate such guidance into requester, recipient, purpose, and action conditions, then resolve conflicts between general and item-specific rules. Ambiguous requests require clarification, conservative action selection, or uncertainty recorded for audit. We leave that parser and conflict resolver open for separate study.

The second problem is abstraction. Many cooperative requests need a summary, constraint, or answer rather than the raw item, turning privacy-preserving disclosure into a generation problem \cite{gao2024memorysharing}. The generator must preserve task-serving content and remove detail outside the stated purpose. The MAGPIE safe-hint examples are small but useful: ``needs stabilization work'' can help planning without exposing crash logs, blame, or customer names. Different requests may require different abstractions of the same item; a release agent, security agent, and work-planning agent should not necessarily receive the same detail.

Abstraction also creates a measurement problem. A disclosure can look safe in isolation while adding to earlier disclosures; several narrow summaries may allow reconstruction of a sensitive fact. This relates to compositional leakage \cite{dwork2008differentialprivacy}, although the setting here is semantic rather than numeric. A MaaS benchmark should measure utility, direct leakage, and leakage under repeated access, while distinguishing harmless detail loss from task-breaking loss. Withholding everything is private but useless; revealing everything is useful but unsafe. The research question is how to measure the middle action without treating it as a vague compromise.

The third problem is accountability. Memory access should leave inspectable records: request tuple $q$, memory item or item class, selected action $a$, policy used, recipient, and time $\tau$. Such records are ordinary in database and security systems, while agent memory systems often treat context as transient prompt material. MaaS treats memory invocation as an event that may need review by a developer, team lead, auditor, or affected user who did not observe the memory flow that shaped an agent decision.

Accountability also includes revocation, time-bounded grants, and interoperability. If a team grants an incident-response agent access to internal notes, that access should end when the incident closes unless renewed. Revocation must address original items, derived abstractions, cached summaries, and downstream copies. This motivates service-layer mediation rather than memory access embedded inside individual agents. Protocols can define tool discovery, messages, or identity, but they do not decide which remembered information flows through a channel. A common MaaS layer lets the memory owner keep one policy surface while multiple agents request bounded access. The same issue appears in regulated public, healthcare, school, and civic workflows: cooperation should depend on the purpose for which information was collected and the form in which it is returned.

For agentic software engineering, this layer should connect to ordinary development artifacts rather than sit beside them. A memory invocation can reference an issue, pull request, test failure, design review, incident, or release gate, making purpose and audit state part of the engineering workflow instead of a hidden prompt decision.

\textbf{Limitations.} MAGPIE is a proxy dataset built from cooperative-agent privacy scenarios, not a log of deployed memory services. The prompt sweep tests framing sensitivity, not the best possible disclosure gate. The regex analysis of safe hints is a lower-bound estimate because it finds only explicit language. We also do not implement a complete MaaS system or evaluate user-facing policy authoring. These limits fit the paper's role as a position paper: define the problem, give a formal handle, and use diagnostic probes to separate memory governance from ordinary retrieval.

\section{Conclusion}
\label{sec:conclusion}

Agent cooperation increasingly relies on memory, yet many designs bind context to the agent, user, project, or platform that collected it. MaaS treats memory invocation as the unit of analysis: who owns the memory, who asks, who receives it, what task it serves, and what purpose justifies access. A mediator can withhold, abstract, or reveal, separating utility, authorization, and disclosure form.

Our diagnostic probes validate the problem boundary: in MAGPIE, relevance weakly separates shareable from private memory (AUROC $0.570$), top-$k$ retrieval exposes private items, prompt framing changes disclosure behavior, and some private items contain safe abstractions. These results support memory services that ask both ``is this relevant?'' and ``is this authorized for this purpose, recipient, and form of disclosure?'' Future work should build mediation mechanisms, policy languages, abstraction tests, and audit tools around that question.

\bibliographystyle{ACM-Reference-Format}
\bibliography{maas}

@misc{lin2026surveysecuritylongtermmemory,
      title={A Survey on the Security of Long-Term Memory in LLM Agents: Toward Mnemonic Sovereignty}, 
      author={Zehao Lin and Chunyu Li and Kai Chen},
      year={2026},
      eprint={2604.16548},
      archivePrefix={arXiv},
      primaryClass={cs.CR},
      url={https://arxiv.org/abs/2604.16548}, 
}

@article{lam2026governing,
  title={Governing evolving memory in {LLM} agents: Risks, mechanisms, and the stability and safety governed memory ({SSGM}) framework},
  author={Lam, Chingkwun and Li, Jiaxin and Zhang, Lingfei and Zhao, Kuo},
  journal={arXiv preprint arXiv:2603.11768},
  year={2026}
}

@inproceedings{
wu2024autogen,
title={AutoGen: Enabling Next-Gen {LLM} Applications via Multi-Agent Conversations},
author={Qingyun Wu and Gagan Bansal and Jieyu Zhang and Yiran Wu and Beibin Li and Erkang Zhu and Li Jiang and Xiaoyun Zhang and Shaokun Zhang and Jiale Liu and Ahmed Hassan Awadallah and Ryen W White and Doug Burger and Chi Wang},
booktitle={First Conference on Language Modeling},
year={2024},
url={https://openreview.net/forum?id=BAakY1hNKS}
}

@inproceedings{
hong2024metagpt,
title={Meta{GPT}: Meta Programming for A Multi-Agent Collaborative Framework},
author={Sirui Hong and Mingchen Zhuge and Jonathan Chen and Xiawu Zheng and Yuheng Cheng and Jinlin Wang and Ceyao Zhang and Zili Wang and Steven Ka Shing Yau and Zijuan Lin and Liyang Zhou and Chenyu Ran and Lingfeng Xiao and Chenglin Wu and J{\"u}rgen Schmidhuber},
booktitle={The Twelfth International Conference on Learning Representations},
year={2024},
url={https://openreview.net/forum?id=VtmBAGCN7o}
}

@misc{anthropic2024mcp,
  title={Model Context Protocol},
  author={{Anthropic}},
  howpublished={\url{https://github.com/modelcontextprotocol }},
  year={2024}
}

@misc{google2025a2a,
  title={Announcing the {Agent2Agent} Protocol ({A2A})},
  author={{Google}},
  howpublished={\url{https://developers.googleblog.com/en/a2a-a-new-era-of-agent-interoperability/ }},
  year={2025}
}

@article{packer2024memgpt,
  publtype={informal},
  author={Charles Packer and Vivian Fang and Shishir G. Patil and Kevin Lin and Sarah Wooders and Joseph E. Gonzalez},
  title={MemGPT: Towards LLMs as Operating Systems},
  year={2023},
  cdate={1672531200000},
  journal={CoRR},
  volume={abs/2310.08560},
  url={https://doi.org/10.48550/arXiv.2310.08560}
}

@inproceedings{zhong2024memorybank,
  title={{MemoryBank}: Enhancing large language models with long-term memory},
  author={Zhong, Wanjun and Guo, Lianghong and Gao, Qiqi and Ye, He and Wang, Yanlin},
  booktitle={Proceedings of the AAAI Conference on Artificial Intelligence},
  volume={38},
  pages={19724--19732},
  year={2024}
}

@inproceedings{xu2025amem,
title={A-Mem: Agentic Memory for {LLM} Agents},
author={Wujiang Xu and Zujie Liang and Kai Mei and Hang Gao and Juntao Tan and Yongfeng Zhang},
booktitle={The Thirty-ninth Annual Conference on Neural Information Processing Systems},
year={2026},
url={https://openreview.net/forum?id=FiM0M8gcct}
}

@article{chhikara2025mem0,
  title={{Mem0}: Building production-ready {AI} agents with scalable long-term memory},
  author={Chhikara, Prateek and Khant, Dev and Aryan, Saket and Singh, Taranjeet and Yadav, Deshraj},
  journal={arXiv preprint arXiv:2504.19413},
  year={2025}
}

@article{rezazadeh2025collaborative,
  title={Collaborative memory: Multi-user memory sharing in {LLM} agents with dynamic access control},
  author={Rezazadeh, Alireza and Li, Zichao and Lou, Ange and Zhao, Yuying and Wei, Wei and Bao, Yujia},
  journal={arXiv preprint arXiv:2505.18279},
  year={2025}
}

@misc{levi2026aip,
      title={AIP: Agent Identity Protocol for Verifiable Delegation Across MCP and A2A}, 
      author={Sunil Prakash},
      year={2026},
      eprint={2603.24775},
      archivePrefix={arXiv},
      primaryClass={cs.CR},
      url={https://arxiv.org/abs/2603.24775}, 
}

@book{nissenbaum2009privacy,
  title={Privacy in context: Technology, policy, and the integrity of social life},
  author={Nissenbaum, Helen},
  year={2009},
  publisher={Stanford University Press}
}

@inproceedings{shao2024privacylens,
  title={PrivacyLens: Evaluating privacy norm awareness of language models in action},
  author={Shao, Yijia and Li, Tianshi and Shi, Weiyan and Liu, Yanchen and Yang, Diyi},
  booktitle={Advances in Neural Information Processing Systems (NeurIPS)},
  year={2024}
}

@misc{shao2025privacylens,
      title={PrivacyLens: Evaluating Privacy Norm Awareness of Language Models in Action}, 
      author={Yijia Shao and Tianshi Li and Weiyan Shi and Yanchen Liu and Diyi Yang},
      year={2025},
      eprint={2409.00138},
      archivePrefix={arXiv},
      primaryClass={cs.CL},
      url={https://arxiv.org/abs/2409.00138}, 
}

@inproceedings{li-etal-2025-1,
    title = "1-2-3 Check: Enhancing Contextual Privacy in {LLM} via Multi-Agent Reasoning",
    author = "Li, Wenkai  and
      Sun, Liwen  and
      Guan, Zhenxiang  and
      Zhou, Xuhui  and
      Sap, Maarten",
    editor = "Derczynski, Leon  and
      Novikova, Jekaterina  and
      Chen, Muhao",
    booktitle = "Proceedings of the The First Workshop on LLM Security (LLMSEC)",
    month = aug,
    year = "2025",
    address = "Vienna, Austria",
    publisher = "Association for Computational Linguistics",
    url = "https://aclanthology.org/2025.llmsec-1.9/",
    pages = "115--128",
    ISBN = "979-8-89176-279-4"
}

@article{cheng2026privact,
  title={Internalizing contextual privacy preservation via multi-agent preference training},
  author={Cheng, Yuhan and Ye, Hancheng and Li, Hai Helen and Sun, Jingwei and Chen, Yiran},
  journal={arXiv preprint arXiv:2602.13840},
  year={2026}
}

@inproceedings{mireshghallah2023confaide,
  author = {Mireshghallah, Niloofar and Kim, Hyunwoo and Zhou, Xuhui and Tsvetkov, Yulia and Sap, Maarten and Shokri, Reza and Choi, Yejin},
  title = {Can LLMs Keep a Secret? Testing Privacy Implications of Language Models via Contextual Integrity Theory},
  booktitle = {International Conference on Learning Representations},
  year = {2024},
  url = {https://proceedings.iclr.cc/paper_files/paper/2024/file/08305d8b2ddab98932c163ea73df065f-Paper-Conference.pdf}
}

@article{juneja2025magpie,
  title={{MAGPIE}: A benchmark for multi-agent contextual privacy evaluation},
  author={Juneja, Gurusha and Pasupulati, Jayanth Naga Sai and Albalak, Alon and Hua, Wenyue and Wang, William Yang},
  journal={arXiv preprint arXiv:2510.15186},
  year={2025}
}

@misc{mem02025openmemory,
  title={{OpenMemory}},
  author={{Mem0}},
  howpublished={\url{https://docs.mem0.ai/openmemory/overview }},
  year={2025}
}

@misc{li2025memosoperatingmemoryaugmentedgeneration,
      title={MemOS: An Operating System for Memory-Augmented Generation (MAG) in Large Language Models}, 
      author={Zhiyu Li and Shichao Song and Hanyu Wang and Simin Niu and Ding Chen and Jiawei Yang and Chenyang Xi and Huayi Lai and Jihao Zhao and Yezhaohui Wang and Junpeng Ren and Zehao Lin and Jiahao Huo and Tianyi Chen and Kai Chen and Kehang Li and Zhiqiang Yin and Qingchen Yu and Bo Tang and Hongkang Yang and Zhi-Qin John Xu and Feiyu Xiong},
      year={2025},
      eprint={2505.22101},
      archivePrefix={arXiv},
      primaryClass={cs.CL},
      url={https://arxiv.org/abs/2505.22101}, 
}

@article{gao2024memorysharing,
  title={Memory sharing for large language model based agents},
  author={Gao, Hang and Zhang, Yongfeng},
  journal={arXiv preprint arXiv:2404.09982},
  year={2024}
}

@inproceedings{dwork2008differentialprivacy,
  title={Differential privacy: A survey of results},
  author={Dwork, Cynthia},
  booktitle={International Conference on Theory and Applications of Models of Computation},
  pages={1--19},
  year={2008}
}

@article{wang2024agentworkflow,
  title={Agent workflow memory},
  author={Wang, Zora Zhiruo and Mao, Jiayuan and Fried, Daniel and Neubig, Graham},
  journal={arXiv preprint arXiv:2409.07429},
  year={2024}
}

@misc{memgraph2025,
  title={MemGraph: Real-time graph streaming platform},
  author={{Memgraph Ltd.}},
  howpublished={\url{https://memgraph.com }},
  year={2025}
}

@misc{steinberger2025openclaw,
  title={OpenClaw: Your own personal {AI} assistant},
  author={Steinberger, Peter},
  howpublished={\url{https://github.com/openclaw/openclaw }},
  year={2025}
}

@misc{nousresearch2026hermes,
  title={Hermes Agent: The agent that grows with you},
  author={{Nous Research}},
  howpublished={\url{https://github.com/nousresearch/hermes-agent }},
  year={2026}
}

@misc{wei2025secondme,
  title={Second Me},
  author={{Second Me}},
  howpublished={\url{https://home.second.me/ }},
  year={2025},
  note={Personal AI identity and memory platform}
}

@misc{nist2025dataasset,
  title={Data asset},
  author={{National Institute of Standards and Technology}},
  howpublished={\url{https://csrc.nist.gov/glossary/term/data_asset }},
  year={2025}
}

@misc{naturalselection2026elys,
  title={Elys: {AI}-native social platform},
  author={{Natural Selection}},
  howpublished={\url{https://apps.apple.com/app/elys }},
  year={2026},
  note={Launched February 2026}
}

@misc{youware2026bloome,
  title={Bloome: {Agentic IM} for multi-bot collaboration},
  author={{YouWare}},
  howpublished={\url{https://bloome.im/welcome }},
  year={2026}
}

@misc{moonshot2026kimi,
  title={Kimi Claw: Complete Guide to Installation, Usage, Pricing, and Advanced Features},
  author={{Kimi-AI.chat}},
  howpublished={\url{https://kimi-ai.chat/guide/kimi-claw/ }},
  year={2026},
  note={Guide describing Kimi Claw and Kimi GroupChat features}
}

@misc{moltbook2026,
  title={MoltBook},
  author={{MoltBook}},
  howpublished={\url{https://www.moltbook.com/ }},
  year={2026},
  note={Agent-native social platform}
}

\end{document}